\documentstyle[11pt]{article}
\normalbaselines
\newcounter{eqnletter}[equation]
\catcode`\@=11
\@addtoreset{equation}{section}   
\catcode`\@=12

\begin{document}

\begin{center}

{\LARGE\bf Quasi-Exactly Solvable Models and W-Algebras}

\vskip 1cm

{\large {\bf A.G. Ushveridze} }\footnote{This work was
partially supported by Deutsche Forschungs Gemeinschaft and Lodz
University Grant N 457.}

\vskip 0.1 cm

Department of Theoretical Physics, University of Lodz,\\
Pomorska 149/153, 90-236 Lodz, Poland\footnote{E-mail
address: alexush@mvii.uni.lodz.pl and alexush@krysia.uni.lodz.pl} \\

\end{center}

\vspace{1 cm}
\begin{abstract}

The relationship between the quasi-exactly solvable problems and 
$W$-algebras is revealed. This relationship enabled one to formulate 
a new general method for building multi-dimensional and multi-channel 
exactly and quasi-exactly solvable models with hermitian hamiltonians. 
The method is based on the use of multi-parameter spectral differential 
equations constructable from generators of finite-dimensional 
representations of simple Lie algebras and from generators of the 
associated $W$-algebras. It is shown that algebras $B_n$, $C_{n+1}$, 
$D_{2n+2}$ with $n>1$ and also algebras $A_1$, $G_2$, $F_4$, $E_7$ 
and $E_8$ always lead to models with hermitian hamiltonians. The situation 
with the remaining algebras $A_n$, $D_{2n+3}$ with $n>1$ and $E_6$ 
is still unclear.

\end{abstract}

\newpage

\section{Introduction}

At present time there are many well elaborated methods
for constructing matrix-valued differential operators
with a complete or partial algebraization of the spectral problem 
(see e.g. the rewiev articles 
\cite{shi(r1), ush(r2), gko(r1), shi(r2), tur(r1)}
and book \cite{ush(b)}).
These operators (which, as a rule, act in the spaces of
polynomials) are non-hermitian and their eigenfunctions are non-normalizable.
If one wants to associate such operators with hamiltonians of some exactly or 
quasi-exactly models of quantum mechanics and relate
their eigenfunctions to wavefunctions, one should find
the appropriate non-unitary transformations ensuring their reduction 
to hermitian form. The problem of constructing such transformations is, 
however, highly non-trivial and admits a regular solution 
only for one-dimensional (non-matrix) second-order 
differential operators \cite{tur(p1), ush(p1)}. A few number of other 
examples related to multi-dimensional and multi-channel
problems can be found in refs. \cite{ush(b), shitur(p1), brikos(p1),
gko(p1)}.

There exists, however, a method in which the hermiticity of the 
resulting operators can be guaranteed from very begining. This is 
the so-called ``inverse method of separation of variables'' proposed 
several years ago in ref. \cite{ush(p2)} 
(see also \cite{ush(b), ush(r1)}) 
and based on the use of systems of 
``multi-parameter spectral equations'', i.e. systems of linear homogeneous 
equations with several spectral parameters. The idea of
the method lies in interpreting the multi-parameter spectral 
equations as equations appearing after the separation of variables in some 
multi-dimensional completely integrable system and in
constructing its integrals of motion by eliminating some of 
spectral parameters. In this case, the eliminable spectral parameters take the
meaning of separation constants.
It is quite obvious that the exact (or quasi-exact)
solvability of the initial multi-parameter spectral
equation implies the exact (or quasi-exact) solvability of
the resulting model. It can also be shown that if the
operators forming the multi-parameter spectral equations 
are hermitian then the integrals of motion of the resulting
multi-dimensional model will also be represented by hermitian operators
(for more details see refs. \cite{ush(r1), ush(b)}).
So we see that the problem of constructing the (separable)
exactly or quasi-exactly solvable models of quantum
mechanics is reduced to the problem of constructing
hermitian exactly or quasi-exactly solvable multi-parameter
spectral equations.

In this paper we show that wide classes of 
exactly and quasi-exactly multi - parameter
spectral equations can easily be constructed from 
genarators of $W$-algebras associated with simple Lie
algebras ${\cal L}$. These classes are parametrized by
the sets of finite-dimensional representations 
$T_\alpha,\ \alpha=1,\ldots, (d-r)/2$ of algebra ${\cal L}$ 
and the sets of rational functions 
$G_i(t),\ i=1,\ldots, r$, where $r$ and $d$ denote the rank
and dimension of algebra ${\cal L}$. The obtained
multi-parameter spectral equations can be
made hermitian for algebras 
$B_n$, $C_{n+1}$, $D_{2n+2}$ 
with $n>1$ and algebras $A_1$, $G_2$, $F_4$, $E_7$ and $E_8$.
This means that $W$-algebras associated with these simple
Lie algebras can be considered as
elementary building blocks for constructing 
wide classes of exactly and quasi-exactly
solvable models with hermitian hamiltonians. 
It is remarkable that these classes
include all the so-called separable quasi-exactly solvable
problems discussed in detail in the book (see e.g. \cite{ush(b)}). 

The fact that the quasi-exactly solvable
models admit a very transparent formulation in terms of $W$-algebras,
has, in our opinion, a grat theoretical importance.
Indeed, remember, that $W$-algebras,
which are the differential polynomial extensions of the Virasoro algebra, and
whose study was first initiated by Zamolodchikov \cite{zam(p1)}, 
play a very important role in numerous branches of mathematical physics. 
For example, they naturally arise in conformal field theories (see e.g.
\cite{fatluk(p1), yam(p1)}), Toda theories
\cite{bilger(p1), bab(p1)}, 
Wess -- Zumino -- Novikov -- Witten
models \cite{bfor(p1)} and Gaudin models \cite{fefrre(p1)}. 
The relation of quasi-exactly solvable models to
Gaudin models and to conformal
field theory has been pointet out many
times \cite{ush(p1), ush(r2), mprst(p1), shi(r2)}. 
The results of our present paper enable one to explain this relationship 
and, what seems especially
important, to extend the list of branches of mathematical
physics to which the quasi-exactly solvable problems are related.

We do not intend to present here any explicit
examples of exactly or quasi-exactly solvable models
obtainable in the framework of the proposed method. This will
be done in an extended version of this paper which will
apear soon in hep-th archive.

\section{Simple Lie algebras}

Let ${\cal L}$ be a simple Lie algebra of rank $r$ and
dimension $d$ and let 
\begin{eqnarray}
{\cal L}={\cal X}\oplus {\cal H}\oplus {\cal Y}
\label{sla.1}
\end{eqnarray}
be its Cartan -- Weyl decomposition. Then in ${\cal L}$ there exist
the elements $X_i\in {\cal X}$, $Y_i \in {\cal Y}$ and
$H_i \in {\cal H}$, $i=1,\ldots, r$ satisfying 
the so-called Weyl commutation relations
\begin{eqnarray}
[H_i,H_k]=0,\quad [H_i,X_k]=+c_{ik}X_k,\quad
[H_i,Y_k]=-c_{ik}Y_k,\quad
[X_i,Y_k]=\delta_{ik}H_i,
\label{sla.2}
\end{eqnarray}
in which $c_{ik},\ i,k=1,\ldots,r$ denote the elements of
Cartan matrix,
\begin{eqnarray}
c_{ik}=\frac{2(\pi_i,\pi_k)}{(\pi_i,\pi_i)},
\label{sla.2a}
\end{eqnarray}
expressed via simple roots $\pi_i,\ i=1,\ldots, r$ of algebra ${\cal L}$.

The elements $H_i,\ i=1,\ldots,r$ form a basis in the
$r$-dimensional Cartan subalgebra ${\cal H}$ of algebra
${\cal L}$ while the elements $X_i,\ Y_i, \ i=1,\ldots,r$ and 
their multiple commutators
generate the $(d-r)/2$-dimensional nilpotent subalgebras ${\cal X}$
and ${\cal Y}$
of algebra ${\cal L}$. All the remaining commutation relations
in algebra ${\cal L}$ can be derived from the basic relations (\ref{sla.2}),
provided that the Cartan matrix is given. 
From these commutation relations it follows that algebra
${\cal L}$ can be considered as graded algebra if one takes by definition 
\begin{eqnarray}
\mbox{deg}\ H_i=0,\qquad \mbox{deg}\
X_i=+ 1,\qquad Y_i=-1.
\label{sla.4}
\end{eqnarray}
Let $N$ be the maximal grading of the elements of algebra
${\cal L}$ and let $d_n,\ n=1,\ldots,N$ 
denote the dimensions of linear spases formed by 
elements of algebra ${\cal L}$ with a given grading $n$.
Then we have 
\begin{eqnarray}
r=d_1\ge d_2 \ge\ldots\ge
d_N=1. 
\label{sla.7}
\end{eqnarray}
Those numbers $n$ for which we have a strict inequality 
$d_n>d_{n+1}$
are called the exponents of algebra ${\cal L}$ and are
usually denoted by $h_1,h_2,\ldots, h_r$. Note that their number is equal
to the rank of algebra ${\cal L}$. From
(\ref{sla.7}) it follows that 
\begin{eqnarray}
h_1=1,\qquad h_r=N
\label{sla.7a}
\end{eqnarray}
for all simple Lie algebras. The exponents $h_1,h_2,\ldots,h_r$ 
determine the orders
\begin{eqnarray}
n_i=h_i+1,\quad i=1,\ldots, r.
\label{sla.8}
\end{eqnarray}
of independent Casimir invariants in algebra ${\cal L}$.
We shall say that a simple Lie algebra is {\it even} if all
its independent Casimir invariants are of an even degree.
The list of even simple Lie algebras contains the following
classical algebras: algebra $A_1$
($n_1=2$), all algebras $B_r$ and $C_r$ ($n_i=2i,\
i=1,\ldots,r$) and
algebras $D_{r}$ with $r=2q$ ($n_i=2i,\ i=1,\ldots,q$
and $n_i=2i-2,\ i=q+1,\ldots,r$). There are also four
exceptional even Lie algebras $G_2$, $F_4$, $E_7$ and $E_8$.
These algebras will play an important role in our further considerations.

Let us now consider two special elements 
\begin{eqnarray}
X=\sum_{i=1}^r X_i,\qquad Y=\sum_{i=1}^r Y_i
\label{sla.9}
\end{eqnarray}
of subalgebras ${\cal X}$ and ${\cal Y}$ and denote by 
${\cal E}\in {\cal X}$ and ${\cal F}\in {\cal Y}$ 
the sets of all elements of ${\cal X}$ and ${\cal Y}$ 
commuting with $X$ and $Y$, respectively:
\begin{eqnarray}
[X,{\cal E}]=0, \qquad [Y,{\cal F}]=0.
\label{sla.10}
\end{eqnarray}
It can be shown that $\dim\ {\cal E}=\dim\ {\cal F}=r$ and the basis elements
$E_i,\ i=1,\ldots,r$ and $F_i,\ i=1,\ldots,r$
in ${\cal E}$ and ${\cal F}$ can be chosen uniquely 
(up to a normalization) if one requires
for them the existence of a certain definite grading. In this case we obtain
\begin{eqnarray}
\mbox{deg}\ E_i=+ h_i,\qquad \mbox{deg}\ F_i=- h_i,
\label{sla.11}
\end{eqnarray}
i.e. the coincidence of gradings of basis elements
with the exponents $h_i$. 
From the definition of elements $E_i$ and $F_i$ it follows that
the linear spaces ${\cal E}$ and ${\cal F}$ form commutative
subalgebras of even Lie algebras ${\cal L}$.
For basis elements $E_1$ and $F_1$ we can take
\begin{eqnarray}
E_1= X, \qquad 
F_1= Y.
\label{sla.12}
\end{eqnarray}
Note that the Weyl commutation relations do not fix uniquely the 
normalization of Weyl generators $X_i$ and $Y_i$. Using this freedom, 
it is not difficult to show that, for any representation of
algebra ${\cal L}$ of a finite dimension $K$, the elements 
$F_i,\ i=1,\ldots,r$ and $E_i,\ i=1,\ldots,r$ can be 
realized by real matrices which are mutually transposed,
$F_i^{\alpha,\beta} = E_i^{\beta,\alpha}, \ \alpha,\beta =
1,\ldots,K$ and symmetric with respect to the second diagonal, 
\begin{eqnarray}
E_i^{\alpha,\beta} = 
E_i^{K+1-\alpha,K+1-\beta}, \qquad F_i^{\alpha,\beta} = 
F_i^{K+1-\alpha,K+1-\beta},\qquad \alpha,\beta=1,\ldots,K.
\label{sla.13}
\end{eqnarray}
Denoting by $J$ the "pseudo unit" matrix with entries
$J^{\alpha,\beta} = \delta_{K+1-\alpha,\beta}$ and using
(\ref{sla.13}) one can claim that the matrices
\begin{eqnarray}
\hat E_i\equiv JE_i,\quad \hat F_i\equiv JF_i
\label{sla.14}
\end{eqnarray}
are real and symmetric, i.e. are hermitian matrices.

In conclusion of this section, let us introduce an
important notion of duality of simple Lie algebras.
Let {\cal L} be a simple Lie algebra of rank $r$ and
dimension $d$ and let $\alpha\in A$ and $\pi_i,\ i=1,\ldots,r$
be its positive and simple roots, respectively.
It is not difficult to check that the vectors 
$\hat\alpha=\alpha/(\alpha,\alpha)\in \hat A$ and 
$\hat \pi_i=\pi_i/(\pi_i,\pi_i),\ i=1,\ldots, r$ also form
the sets of positive and simple roots of a certain simple
Lie algebra $\hat {\cal L}$ of the same rank $r$ and
dimension $d$. We shall call the algebras ${\cal L}$ and $\hat {\cal L}$
dual. The Cartan matrices $c_{ik}$ and $\hat c_{ik}$ of 
algebras ${\cal L}$ and $\hat {\cal L}$ are related to each
other by a transposition: $\hat c_{ik}= c_{ki}$. For this reason,
those of simple Lie algebras whose Cartan matrices are
symmetric are self-dual ${\cal L}=\hat {\cal L}$. 
These are two series of classical algebras
$A_r$, $D_r$ and three exceptional algebras $E_6$, $E_7$ and
$E_8$. The two remaining exceptional algebras $G_2$ and $F_4$ are
also self-dual although their Cartan matrices are
non-symmetric. In this case the duality transformation
changes the numeration of vertices of the corresponding
Dynkin schemes. The Cartan matrices of 
two remaining series of classical algebras $B_r$
and $C_r$ are mutually transposed and therefore algebras
$C_r$ and $B_r$ are mutually dual.

\section{Drinfeld -- Sokolov theorems}

Denote by $\Phi_{\cal H}$ and $\Phi_{\cal Y}$ the spaces of
analytic functions of a complex parameter $t$ taking their walues in
algebras ${\cal H}$ and ${\cal Y}$, respectively. Introduce
also the following differential operator
\begin{eqnarray}
D=\imath\frac{\partial}{\partial t}+X
\label{dsc.1}
\end{eqnarray}
acting in a direct product of the space of analytic
functions of $t$ and the representation space of
algebra ${\cal L}$.

\medskip
{\bf Theorem 1.} For any functions $H(t)\in \Phi_{\cal H}$ and
$Y(t)\in \Phi_{\cal Y}$ and any $\Theta(t)\in \Phi_{\cal Y}$ 
the folowing equality holds
\begin{eqnarray}
e^{-\Theta(t)}\left\{
D+ H(t)+Y(t)\right\}e^{+\Theta(t)}=
D+ \bar H(t)+\bar Y(t)
\label{dsc.2}
\end{eqnarray}
with certain $\bar H(t)\in \Phi_{\cal H}$ and
$\bar Y(t)\in \Phi_{\cal Y}$.

\medskip
{\bf Theorem 2.} For any functions $H(t)\in \Phi_{\cal H}$ and
$Y(t)\in \Phi_{\cal Y}$ there exists an unique exactly computable
function $\Theta(t)\in \Phi_{\cal Y}$ for which
\begin{eqnarray}
e^{-\Theta(t)}\left\{
D+ H(t)+Y(t)\right\}e^{+\Theta(t)}=
D+ \sum_{i=1}^r F_i\ U_i[H(t),Y(t)]
\label{dsc.3}
\end{eqnarray}
where $U_i[H(t),Y(t)]
\ i=1,\ldots,r$ are certain differential polynomials in
components of functions $H(t)$ and $Y(t)$.

\medskip
These two theorems have been formulated and proven by Drinfeld and
Sokolov in their review paper \cite{drisok(p1)}. 

\medskip
{\bf Theorem 3.} For any set of functions $z_i(t),\ i=1,\ldots,r$ 
there exists an unique exactly computable 
function $\Theta(t)\in \Phi_{\cal Y}$ such that 
\begin{eqnarray}
e^{-\Theta(t)}\left\{
D-\imath\sum_{i=1}^r H_i z_i(t)
\right\}e^{+\Theta(t)}=
D+ \sum_{i=1}^r(-\imath)^{n_i}F_i\ W_i[z_1(t),\ldots, z_r(t)]
\label{dsc.4}
\end{eqnarray}
where $W_i[z_1(t),\ldots,z_r(t)],
\ i=1,\ldots,r$ are certain real differential polynomials
in functions $z_i(t),\ i=1,\ldots,r$ and $n_i,\ i=1,\ldots,r$ are
the orders of the independent Casimir invariants. 
If the analytic functions $z_i(t),\ i=1,\ldots,r$ and
the operator $\partial/\partial t$ are considered as graded objects
with unit gradings
\begin{eqnarray}
\mbox{deg}\ z_i(t)=1,\qquad \mbox{deg}\ \partial=1,
\label{dsc.5}
\end{eqnarray}
then $W_i[z_1(t),\ldots,z_r(t)],
\ i=1,\ldots,r$ become homogeneous (graded) differential polynomials
of degrees
\begin{eqnarray}
\mbox{deg}\ W_i[z_1(t),\ldots, z_r(t)]=n_i,
\qquad i=1,\ldots, r.
\label{dsc.6}
\end{eqnarray}

\medskip
The first part of this theorem (which is a slight
modification of assertions proved in refs. \cite{drisok(p1), bfor(p1)}) 
is a trivial corrolary of 
the second Drinfeld -- Sokolov theorem. The second part follows from
the observation that if the left-hand side of (\ref{dsc.4})
has a definite (unit) grading, then the right-hand
side should have the same grading. In particular, we should
have the equalities
\begin{eqnarray}
\mbox{deg}\ F_i + \mbox{deg}\ W_i[z_1(t),\ldots, z_r(t)]=1,
\qquad i=1,\ldots, r,
\label{dsc.7}
\end{eqnarray}
which, after taking into account formula (\ref{sla.11})
lead to (\ref{dsc.6}). 

Thus, we see that any simple Lie algebra ${\cal L}$
of rank $r$ generates $r$ homogeneous differential polynomials 
$W_i[z_1(t),\ldots,z_r(t)],\ i=1,\ldots,r$ of degrees
$n_i=h_i+1,\ i=1,\ldots,r$ coinciding with the orders of
independent Casimir invariants in algebra ${\cal L}$. We
know from ref. \cite{bfor(p1)} that these polynomials form
a basis in $W$-algebra associated with simple Lie
algebra ${\cal L}$. An explicit form of these generators
for various simple Lie algebras can be found 
in refs. \cite{bfor(p1), ush(p3)}. Following ref. \cite{ush(p3)},
we shall call these generators ``Riccatians''
because, as we see below, the simplest polynomial of this
sort associated with algebra $A_1$ is the ordinary Riccati polynomial
$W_1[z(t)]=z'(t)+z^2(t)$ entering into Riccati equation.

\section{The analytic properties of Riccatians}

One of the most interesting properties of Riccatians
is that they may be regular
functions of $t$ even if the functions 
$z_1(t),\ldots,z_r(t)$ are singular. 

\medskip
{\bf Theorem 4.} Let functions $z_i(t),\ i=1,\ldots,
r$ have the following form
\begin{eqnarray}
z_i(t)=\eta_i(t)+\frac{\delta_{ij}}{t-\xi},\qquad
i=1,\ldots,r
\label{dsc.8}
\end{eqnarray}
where $j$ is an arbitrarily fixed natural number $1\le j\le r$,
$\xi$ is an arbitrary complex number and 
$\eta_i(t)$ are arbitrary functions satisfying the only condition
\begin{eqnarray}
\sum_{i=1}^r c_{ij} \eta_i(\xi)=0. 
\label{dsc.9}
\end{eqnarray}
Then the differential polynomials $W_i[z_1(t),\ldots,z_r(t)]$
considered as analytic functions of $t$ are regular
at the point $t=\xi$.

\medskip
The proof of this theorem we divide into two parts. First of
all, let us substitute 
\begin{eqnarray}
H(t)=-\imath\sum_{i=1}^r H_i z_i(t)=
-\imath\sum_{i=1}^r H_i \eta_i(t)-\imath
\frac{H_j}{t-\xi}, \qquad Y(t)=0
\label{dsc.10}
\end{eqnarray}
into the left-hand side of formula (\ref{dsc.2}) and, taking
\begin{eqnarray}
\Theta(t)=\theta(t)Y_j
\label{dsc.11}
\end{eqnarray}
and using commutation relations (\ref{sla.2}),
try to compute the right-hand side of (\ref{dsc.2}).
Remembering that $c_{jj}=2$, we obtain
\begin{eqnarray}
&~&e^{-\Theta(t)}\left\{
D+ H(t)+Y(t)\right\}e^{+\Theta(t)}=\nonumber\\
&~&\nonumber\\
&=&e^{-\theta(t)Y_j}\left\{
\imath\frac{\partial}{\partial t}+\sum_{i=1}^r X_i-\imath
\sum_{i=1}^r H_i\eta_i(t)-\imath\frac{H_j}{t-\xi}\right\}
e^{+\theta(t)Y_j}=\nonumber\\
&~&\nonumber\\
&=&\left(\imath\frac{\partial}{\partial t}+\sum_{i=1}^r X_i-\imath
\sum_{i=1}^r H_i\eta_i(t)-
\imath\frac{H_j}{t-\xi}\right)+\nonumber\\
&+&\left[\theta(t)Y_j,
\left(\imath\frac{\partial}{\partial t}+\sum_{i=1}^r X_i-\imath
\sum_{i=1}^r H_i\eta_i(t)-\imath\frac{H_j}{t-\xi}\right)\right]
+\nonumber\\
&+\frac{1}{2}&\left[\theta(t)Y_j,\left[\theta(t)Y_j,
\left(\imath\frac{\partial}{\partial t}+\sum_{i=1}^r X_i-\imath
\sum_{i=1}^r H_i\eta_i(t)-\imath\frac{H_j}{t-\xi}\right)
\right]\right]=\nonumber\\
&~&\nonumber\\
&=&\imath\frac{\partial}{\partial t}+\sum_{i=1}^r X_i-\imath
\sum_{i=1}^r H_i\eta_i(t)-H_j\left(\frac{\imath}{t-\xi}+
\theta(t)\right)+\nonumber\\
&+&Y_j\left(-\imath\frac{\partial \theta(t)}{\partial t}-\imath
\frac{2\theta(t)}{t-\xi}-\theta^2(t)-\imath
\sum_{i=1}^r c_{ij}\eta_i(t)\theta(t)
\right)=\nonumber\\
&~&\nonumber\\
&=&D+ \bar H(t)+\bar Y(t)
\label{dsc.12}
\end{eqnarray}
with
\begin{eqnarray}
\bar H(t)&=&
-\imath\sum_{i=1}^r H_i\eta_i(t)-H_j\left(\frac{\imath}{t-\xi}+
\theta(t)\right),\nonumber\\
\bar Y(t)&=&
Y_j\left(-\imath\frac{\partial \theta(t)}{\partial t}-\imath
\frac{2\theta(t)}{t-\xi}-\theta^2(t)-\imath
\sum_{i=1}^r c_{ij}\eta_i(t)\theta(t)
\right)
\label{dsc.13}
\end{eqnarray}
From (\ref{dsc.13}) it follows that if one takes in (\ref{dsc.13})
\begin{eqnarray}
\theta(t)=-\frac{\imath}{t-\xi},
\label{dsc.14}
\end{eqnarray}
then many singular terms in the expressions for $\bar H(t)$ 
and $\bar Y(t)$ cancel and
they take the form
\begin{eqnarray}
\bar H(t)=-\imath\sum_{i=1}^r H_i\eta_i(t),\qquad
\bar Y(t)=-\imath 
Y_j\sum_{i=1}^r c_{ij}\frac{\eta_i(t)}{t-\xi}
\label{dsc.15}
\end{eqnarray}
It is also not difficult to see that both the functions $\bar H(t)$ 
and $\bar Y(t)$ are regular at the point $t=\xi$ if 
the conditions (\ref{dsc.9}) are satisfied.

The second part of the proof is based on the use of theorem
2, according to which, there exist an unique 
function $\bar\Theta(t)$ transforming
the operator $D+\bar H(t)+\bar Y(t)$ into the
canonical form $D+\sum_{i=1}^r F_i 
\bar W_i[\bar H(t),\bar Y(t)]$. 
Due to the uniqueness of the canonical form, 
functions $\bar W_i[\bar H(t),\bar Y(t)]$
coincide with function
$(-\imath)^{n_i}W_i[z_1(t),\ldots,z_r(t)]$. Since functions 
$\bar W_i[\bar H(t),\bar Y(t)]$ are  differential polynomials 
in components of functions $\bar H(t)$ and $\bar
Y(t)$ which are regular
at the point $t=\xi$, they also should be regular at
this point. This fact, in turn, implies the regularity of functions
$W_i[z_1(t),\ldots,z_r(t)]$ at $t=\xi$ and
completes the proof of theorem 4.

Thus, we can conclude that Riccatians
have a remarkable property, according to which,
for any $j=1,\ldots,r$ and any
$z_1(t),\ldots,z_r(t)$, functions
\begin{eqnarray}
\omega_i(t)=
W_i\left[z_1(t),\ldots,z_{j-1}(t),
z_j(t)+\frac{1}{t-\xi},z_{j+1}(t),
\ldots,z_r(t)\right]
\label{dsc.16}
\end{eqnarray}
are regular at the point $t=\xi$ provided that the
condition (\ref{dsc.9}) is satisfied. This condition will
play a very important role in our further considerations.
It is convenient to rewrite it in the form
\begin{eqnarray}
\sum_{k=1}^r (\hat\pi_j,\hat\pi_k) z_k(\xi)=0,
\label{dsc.17}
\end{eqnarray}
i.e., in terms of simple roots of algebra $\hat{\cal L}$
dual to algebra ${\cal L}$.

\section{Generalized Riccati equations.}

Let $W_i[z_1(t),\ldots,z_r(t)],\ i=1,\ldots, r$
be Riccatians associated with a certain simple Lie algebra
${\cal L}$ of rank $r$.
Consider a system of the followig formal relations
\begin{eqnarray}
W_i[z_1(t),\ldots,z_r(t)]=\omega_i(t),\qquad
i=1,\ldots, r
\label{re.1}
\end{eqnarray}
in which $z_1(t), \ldots, z_r(t)$ and 
$\omega_1(t), \ldots, \omega_r(t)$ are assumed to
be some analytic functions of a complex variable $t$.

First of all note that the relations (\ref{re.1}) (if one
considers them as equations) admit at
least two interpretations:

1. {\it Function $z_1(t), \ldots, z_r(t)$ are given, 
and functions $\omega_1(t), \ldots, \omega_r(t)$
are being sought}. This problem is trivial
and has a unique solution.

2. {\it Functions $\omega_1(t), \ldots, \omega_r(t)$ are given, 
and functions $z_1(t), \ldots, z_r(t)$ are being sought}. 
This problem is nothing else
than a generalization of the ordinary Riccati equation. 
Except some very special cases this problem cannot be solved in 
quadratures even in the simplest case of algebra $A_1$.

It turns out however that along with these two polar 
interpretations, there exists an
interesting intermediate one which leads
to a very rich set of solutions and has a great theoretical
importance.  Roughly speaking, the idea of this
interpretation is to fix appropriately {\it some parts} 
of {\it both} sets of functions
$z_1(t), \ldots, z_r(t)$ and 
$\omega_1(t), \ldots, \omega_r(t)$ and state the problem of
finding the {\it remaining parts} of these functions. Before
giving a rigorous formulation of this problem, let us
introduce some necessary notions and notations.

Let ${\cal R}$ be the class of all $r$-component vector 
rational functions $r(t)$
of a single complex variable $t$. This class can obviously be 
viewed as an infinite-dimensional linear vector space with
a basis consisting of the {\it elementary
rational functions}. Let $A$ be the a finite set of nonequal 
fixed complex numbers (one of
which may be the infinity), and
$B$ be its complement. Denote by 
${\cal R}_A$ and ${\cal R}_B$ the classes of those
$r$-component vector rational 
functions from ${\cal R}$
whose singularities belong only to the sets $A$ and $B$, respectively. 
Considering the classes ${\cal R}_A$ and ${\cal R}_B$ as
linear vector spaces we can write ${\cal R}_A\oplus{\cal R}_B={\cal R}$.
Denote also by ${\cal P}_A$ and ${\cal P}_B$ the projectors from ${\cal R}$ 
onto ${\cal R}_A$ and ${\cal R}_B$ for which we obviously
have ${\cal P}_A+{\cal P}_B=1$. Let us now formulate the problem 
of our interest.

\medskip
{\bf Problem.} Find functions 
$z_1(t), \ldots, z_r(t) \in {\cal R}$
and $\omega_1(t), \ldots, \omega_r(t) \in {\cal R}$
satisfying the relation (\ref{re.1}) under 
two additional constrains:
\begin{eqnarray}
{\cal P}_A\ z_i(t)=G_i(t), \quad {\cal P}_B\ \omega_i(t)=0,
\quad i=1,\ldots, r
\label{re.2}
\end{eqnarray}
in which $G_i(t)\in {\cal R}_A,\ i=1,\ldots, r$ are given functions.

\medskip
The first constrain means that all the functions
$z_i(t),\ i=1,\ldots, r$
should have some fixed projections $G_i(t),\
i=1,\ldots, r$ onto the space ${\cal R}_A$ and, in principle, 
may have arbitrary singularities outside the set $A$.
The second constrain means however that not any singularities of
functions $z_i(t), \ i=1,\ldots, r$ lying outside the set $A$ are
admissible, but only those, at which the functions
$\omega_i(t), \ i=1,\ldots, r$ are regular. 

Now we are ready to formulate the following theorem.

\medskip
{\bf Theorem 5.} The solution of problem 1 has
the following form:
\begin{eqnarray}
z_i(t) = G_i(t)+
\sum_{j=1}^{M_i}\frac{1}
{t - \xi_{ij}},  
\label{re.3a}
\end{eqnarray}
\begin{eqnarray}
\omega_i(t) = W_i
\left[G_1(t)+ \sum_{j=1}^{M_1}\frac{1}
{t - \xi_{1j}},\ldots,G_r(t)+ \sum_{j=1}^{M_r}\frac{1}
{t - \xi_{rj}} \right],
\label{re.3b}
\end{eqnarray}
where $M_i, \ i=1,\ldots, r$ are arbitrary non-negative
integers, and the parameters $\xi_{ij}, \ j=1,\ldots, M_i,
\ i=1,\ldots, r$ satisfy the system of equations
\begin{eqnarray}
G_i(\xi_{ij})+
\sum_{k}\sum_{l=1}^{M_{k}}\frac{(\hat\pi_i,\hat\pi_{k})}
{\xi_{ij} - \xi_{kl}}=0, \quad j=1,\ldots, M_i,
\ i=1,\ldots, r.
\label{re.4}
\end{eqnarray}
For any set of numbers $M_1,\ldots, M_r$ the
equations (\ref{re.4}) have a finite set of solutions and
therefore, the spectrum of the the system of Riccatians
is infinite and discrete.

\medskip
{\bf Proof.} The proof of this theorem is very simple.
Rewritting (\ref{re.3a}) in
one of the following $M_1+\ldots+M_r$ forms
\begin{eqnarray}
z_i(t) = \eta_{ij}(t)+ \frac{\delta_{ij}}{t -
\xi_{ij}}, \quad j=1,\ldots, M_i, \quad i=1,\ldots, r,
\label{re.6}
\end{eqnarray}
where 
\begin{eqnarray}
\eta_{ij}(t)=
G_i(t)+\sum_{j'\neq j}^{M_i}\frac{1}{t-\xi_{ij'}}
\label{re.7}
\end{eqnarray}
is a function regular at $t=\xi_{ij}$,
and using formula (\ref{dsc.17}), we
can conclude that the conditions of regularity of 
functions $\omega_i(t)]$ at the points $t=\xi_{ij}$ are
given by formula (\ref{re.4}).
This completes the proof of the theorem.

\section{Riccatians and linear differential equations}

Assume that generators $X_i=X_i[\Lambda]$, $Y_i=Y_i[\Lambda]$ and 
$H_i=H_i[\Lambda]$ are 
matrices realizing a certain finite-dimensional
representation $T_{\Lambda}$ of algebra ${\cal L}$ characterized by
highest weight $\Lambda$. The corresponding highest weight
vector we denote by $|\Lambda\rangle$. The 
operators $\hat D$ and matrices 
$\hat F_i$ we denote by $\hat D[\Lambda]$ and $\hat F_i[\Lambda]$. 

Let $W_i[z_1(t),\ldots,z_r(t)],\ i=1,\ldots,r$ be the
Riccatians associated with algebra ${\cal L}$.
They, obviously, do not depend on the sort of the representation
because for their construction it is sufficient to use the
commutation relations in algebra ${\cal L}$.
Consider the equation
\begin{eqnarray}
\left(D[\Lambda]+\sum_{i=1}^r(-\imath)^{n_i} F_i[\Lambda]\cdot W_i[z_1(t),
\ldots, z_r(t)]\right)
\phi[\Lambda,t]=0
\label{cor.1}
\end{eqnarray}
for a vector function $\phi[\Lambda,t]$.
The solutions of this equation do, obviously, depend  on
the representation $T_\Lambda$.
The number of these solutions is equal to the dimension 
of the representation $T_{\Lambda}$.
Below we demonstrate that at least one of its solutions can
be constructed explicitly.
To do this, let us rewrite (\ref{cor.1}) in the form
\begin{eqnarray}
\left(
D[\Lambda]-\imath\sum_{i=1}^r H_i[\Lambda] z_i(t)
\right)\psi[\Lambda,t]=0,
\label{cor.4}
\end{eqnarray}
Here $\psi[\Lambda,t]$ is another vector function related to
$\phi[\Lambda,t]$ by the formula
\begin{eqnarray}
\phi[\Lambda]=\exp\left(\Theta[\Lambda,t]\right) \psi[\Lambda,t].
\label{cor.5}
\end{eqnarray}
where $\Theta[\Lambda,t]$ is an explicitly constructable
matrix function. This follows from the proof of Drinfeld --
Sokolov theorems (for more details see ref. \cite{drisok(p1), bfor(p1)}).
Formula (\ref{cor.5}) enables
one to construct explicit solutions of the equation (\ref{cor.1}).
In order to demonstrate this,
let us consider the following ansatz for vector function $\psi[\Lambda,t]$ 
of the form
\begin{eqnarray}
\psi[\Lambda,t]=\upsilon[\Lambda,t]|\Lambda\rangle
\label{cor.4b}
\end{eqnarray}
and note that, in this case,
\begin{eqnarray}
X_i[\Lambda]\psi[\Lambda,t]=0, \quad H_i[\Lambda]\psi[\Lambda,t]=
\Lambda_i\psi[\Lambda,t],
\qquad i=1,\ldots,r
\label{cor.8}
\end{eqnarray}
where $\Lambda_i,\ i=1,\ldots,r$ are the comonents of
highest weight $\Lambda$. From (\ref{cor.8}) it follows that function
$\psi[\Lambda,t]$ satisfies the equation (\ref{cor.4}),
provided that $\upsilon[\Lambda,t]$ is a solution of the equation
\begin{eqnarray}
\left(\frac{\partial}{\partial t}-\sum_{i=1}^r 
\Lambda_i z_i(t)\right)\upsilon[\Lambda,t]=0.
\label{cor.9}
\end{eqnarray}
Solving (\ref{cor.9}), and substituting the result into
(\ref{cor.4b}) and (\ref{cor.5}), we obtain
\begin{eqnarray}
\phi[\Lambda,t]=\prod_{i=1}^r \exp\left(\Lambda_i\int z_i(t)dt\right)
\exp\left(\Theta[\Lambda,t]\right) |\Lambda\rangle.
\label{cor.10}
\end{eqnarray}
So we have shown that the equation
(\ref{cor.1}) is exactly solvable for any set of functions
$z_i(t), \ i=1,\ldots, r$. 

For further convenience, it is, however, 
reasonable to rewrite equation (\ref{cor.1}) in different form.
Remember that the matrices $F_i[\Lambda], \ i=1,\ldots,r$ are not
hermitian, while the matrices $\hat F_i[\Lambda]=J[\Lambda]F_i[\Lambda], \
i=1,\ldots,r$ (in which $J[\Lambda]$ denotes the "pseudo
unit" operator in the representation space)
are (see section 2). For the same reason, the operator $\hat D[\Lambda]=
J[\Lambda]D[\Lambda]$
is a hermitian matrix differential operator. Multiplying
the equation (\ref{cor.1}) by $J[\Lambda]$ we obtain a new exactly
solvable equation having the same solutions as (\ref{cor.1}). It is easily
seen that the linear differential operator $\hat D[\Lambda]+\sum_{i=1}^r
(-\imath)^{n_i} \hat F_i[\Lambda]\cdot W_i$
characterizing this equation may be hermitian only
for even simple Lie algebras, i.e. algebras with even Casimir invariants.
For these algebras the equation under consideration takes
the form
\begin{eqnarray}
\left(\hat D[\Lambda]-\sum_{i=1}^r \hat F_i[\Lambda]\cdot 
W_i[z_1(t),\ldots,z_r(t)]\right)
\phi[\Lambda,t]=0
\label{cor.12}
\end{eqnarray}
and, as we show in next section, can be used for building
hermitian multi-parameter spectral equations.

\section{Riccatians and multi-parameter spectral equations}

Up to now functions $z_1(t),\ldots,z_r(t)$ were assumed to
be arbitrary analytic functions of $t$. Let us now assume that
these functions are solutions of the system of generalized 
Riccati equations (\ref{re.1}) and (\ref{re.2}) for a given
set of characteristic functions $G_i(t)$.
For the sake of simplicity we can assume that $G_i(t)$ are
non-degenerate real rational functions having only the simple poles
at $K$ points $a_1,\ldots,a_K$ of the real $t$-axis.
We can represented these functions in the form
\begin{eqnarray}
G_i(t)=\sum_{k=1}^K\frac{g_{ik}}{t-a_k},\qquad i=1,\ldots,r
\label{m.1}
\end{eqnarray}
where $g_{ik}$ are certain real residues.

According to the results of section 5, the ``functional values'' 
of Riccatians for the solutions of the generalized Riccati equation are
\begin{eqnarray}
W_i[z_1(t),\ldots,z_r(t)]=\frac{\sum_{n=0}^{(K-1)n_i} q_{i,n} t^n}
{\left[\prod_{k=1}^K (t-a_k)\right]^{n_i}}
\label{m.2}
\end{eqnarray}
$q_{i,n}, \ n=0,\ldots,(K-1)n_i$ are some coefficients in which all the
dependence on the sort of solution (i.e. on the numbers
$\xi_{ij}$) is concentrated.
Analysing the behaviour of these functions near singular
points, we can see that, for any given $i$, the number of
independent coefficients $q_{i,n}$ in (\ref{m.2}) is $N_i=(K-1)h_i$
where $h_i=n_i-1$ is an exponent of algebra
${\cal L}$. Using the expansions (\ref{m.2}), we can write
\begin{eqnarray}
\sum_{i=1}^r \hat F_i[\Lambda]\cdot W_i[z_1(t),\ldots,z_r(t)] 
= \hat\Omega[\Lambda,t]+\sum_{j=1}^N\hat \Omega^j[\Lambda,t] 
\lambda_j
\label{m.3}
\end{eqnarray}
where $\hat \Omega[\Lambda,t]$  and $\hat \Omega^j[\Lambda,t]$ 
are some fixed hermitian and rational matrix valued functions
and  $\lambda_j$ are some linear combinations of independent numbers 
$q_{i,n}$. The total number of parameters $\lambda_j$ is
$N=N_1+\ldots+N_r = (K-1)(h_1+\ldots+h_r) = (K-1)A$, where
$A=(d-r)/2$ is the number of all positive roots in algebra
${\cal L}$.
Introducing the notation
\begin{eqnarray}
\hat D[\Lambda,t]=\hat D[\Lambda]-\hat\Omega[\Lambda,t]
\label{m.3a}
\end{eqnarray}
and using (\ref{m.3}) and (\ref{cor.12}), we arrive at the equation
\begin{eqnarray}
\left(\hat D[\Lambda,t]-\sum_{j=1}^N\hat \Omega^j[\Lambda,t]\lambda_j\right)
\phi[\Lambda,t]=0
\label{m.4}
\end{eqnarray}
which, for some special values of parameters $\lambda_j$, admits 
explicit solutions 
\begin{eqnarray}
\phi[\Lambda,t]= \prod_{i=1}^r\prod_{j=1}^{M_i}(t-\xi_{ij})^{\Lambda_i}
\exp\left(\Lambda_i\int G_i(t)dt\right)
\exp\left(\Theta[\Lambda,t]\right) |\Lambda\rangle.
\label{m.5}
\end{eqnarray}

It turns out that in order to construct the system 
of exactly solvable and hermitian multi-parameter spectral equations,
it is sufficient to take $A$ equations
(\ref{m.4}) associated with $A$ different representations 
of algebra ${\cal L}$. In order to show this, let us first 
introduce two discrete functions
\begin{eqnarray}
\alpha_i=1+\left[\frac{i-1}{K-1}\right],\qquad
\beta_i=1+(K-1)\left\{\frac{i-1}{K-1}\right\},
\qquad i=1,\ldots, N
\label{m.6}
\end{eqnarray}
in which $[\ ]$ denotes the integer part and $\{\ \}$ is the fraction
part of a rational number. Note that function $\alpha_i$
is non-decreasing and takes $K-1$ times each integer value 
from $1$ to $A$. As to the function $\beta_i$, it consists of $A$ pieces
monotonically increasing from $1$ to $K-1$.

Now consider $A$ arbitrarily chosen representations of
algebra ${\cal L}$ with different highest weights $\Lambda^\alpha, \
\alpha=1,\ldots, A$ and dimensions $d_\alpha,\ \alpha=1,\ldots,A$.
Introduce the spaces $V_i, \ i=1,\ldots, N$ by the formula
\begin{eqnarray}
V_i=V_{\beta_i}[\Lambda^{\alpha_i}], \qquad i=1,\ldots, N
\label{m.7}
\end{eqnarray}
where $V_\beta[\Lambda^\alpha]$ denotes the space of all
$d_\alpha$-dimensional vector analytic functions vanishing sufficiently 
rapidly at the ends of the interval $[a_\beta,a_{\beta+1}]$
and endowed by the scalar products $\langle\ \phi,\phi'
\rangle = \int_{a_\beta}^{a_{\beta+1}}\phi^+(t)\phi'(t)dt$.
Note that the operators $\hat D[\Lambda^\alpha,t]$ and 
$\hat \Omega_j[\Lambda^\alpha,t]$
are hermitian in each of the spaces $V_\beta[\Lambda^\alpha],\ 
\beta=1,\ldots,K-1$. Moreover, the explicit solutions 
$\phi[\Lambda^{\alpha},t]$ given by formula (\ref{m.5})
belong to the spaces $V_\beta[\Lambda^\alpha],\ 
\beta=1,\ldots,K-1$ for sufficiently large positive values
of residues $g_{ik}$ (see formula (\ref{m.1})).
Using these facts and introducing the following 
new notations for these operators
and functions
\begin{eqnarray}
L_i=\hat D[\Lambda^{\alpha_i},t]\in \mbox{End}\ [V_i],\qquad
i=1,\ldots, N
\label{m.8}
\end{eqnarray}
\begin{eqnarray}
L_i^j=\hat \Omega^j[\Lambda^{\alpha_i},t]\in \mbox{End}\ [V_i],\qquad
i=1,\ldots, N,\quad j=1,\ldots, N,
\label{m.9}
\end{eqnarray}
\begin{eqnarray}
\phi_i=\phi[\Lambda^{\alpha_i},t]\in V_i
\label{m.10}
\end{eqnarray}
we arrive at the system of exactly solvable hermitian multi-parameter
spectral equations 
\begin{eqnarray}
V_i\ni\left[L_i-\sum_{j=1}^N L_i^j
\lambda_i\right]\phi_i=0,\qquad V_i\ni\phi_i\neq 0 \qquad i=1,\ldots,N
\label{m.11}
\end{eqnarray}
whose form exactly coincides with that discussed in sections 2 -- 6.

\section{Exact and quasi-exact solvability}

The exact solvability and hermiticity of system (\ref{m.11}) means
that it can be reduced to exactly solvable and completely
integrable quantum models realized
by hermitian, in general, matrix hamiltonians acting in the spaces of, in
general, multi-component wavefunctions. Beyond any doubt,
the class of such models should be closely related to the
famous Gaudin models because of the coincidence of their
Bethe ansatz solutions with those of the 
generalized Riccati equations.

It turns out that the same equations can be used for
building the quasi-exactly solvable problems. In order to
show this, it is sufficient to prove the existence of spectral
parameters (or their linear combinations) with degenerate spectra.
Fortunately, the proof is trivial. Indeed, assume that functions 
$z_1(t),\ldots,z_r(t)$ satisfy the generalized Riccati equation with 
functions $G_i(t)$ given by formula (\ref{m.1}). Consider the expansion 
(\ref{m.2}), multiply it by $t^{n_i}$ and take the limit 
$t\rightarrow \infty$. From formula (\ref{re.3b}) it follows that the leading
spectral parameter $q_{(K-1)n_i}$ depends only on the residues 
$g_{ik}$ and numbers $M_1,\ldots,M_r$. Since this holds for
any $i=1,\ldots, r$, we can conclude that the set of
spectral parameters of equation (\ref{m.11}) contains $r$ parameters 
with degenerate spectra. 

According to the general theory of multi-parameter spectral equations
given in refs. \cite{ush(r1), ush(b)}, the
hamiltonians of quasi-exactly solvable models associated with
equations (\ref{m.11}) are hermitian and the corresponding
wavefunctions are normalizable. Note that one- and
multi-dimensional quasi-exactly
solvable models corresponding to algebra $A_1$ (with a single
Riccatian $W[z(t)]=z'(t)+z^2(t)$) have already been studied in
detail in refs. \cite{ush(p1), ush(p2), ush(r1), ush(b)}.
The next interesting case is that of algebra $B_2$ (because
the algebra $A_2$ does not lead to hermitian hamiltonians). An
explicit construction of the corresponding quasi-exactly
solvable models is an interesting mathematical problem
which we intend to consider in next publication.

\section{Acknowledgements}

I am grateful to E. Frenkel who gave me an exhaustible list of references 
to $W$-algebras and Drinfeld -- Sokolov reduction scheme.

\end{document}